\documentclass{elsart}

\usepackage[]{graphicx}
\usepackage{amssymb}

\begin{document}

\begin{frontmatter}

\title{The pseudogaps in multiband electron-doped cuprate superconductivity}

\author{N.~Kristoffel$^\dagger$\thanksref{mail}},
\author{P.~Rubin$^\dagger$}

\address{$^\dagger$Institute of Physics, University of Tartu,
Riia 142, 51014 Tartu, Estonia}
\thanks[mail]{Corresponding author:
E-mail: kolja@fi.tartu.ee}

\begin{abstract}

The pseudogap (PG) excitations in the framework of multiband superconducti\-vity are analysed.
 Calculations for the doping phase diagram of electron-doped cuprate superconductors have been made.
 A nonrigid multiband model has been used.
The interband pairing acts
between the upper Hubbard band (UHB) and midgap states created (and shifted
upwards) by doping. The appearance of two PG and two phase diagram critical
points (CP) is suggested. The larger antinodal PG reflects the destruction
of the long-range antiferromagnetic (AF) order.  The second, smaller PG is a
near-nodal event with exposed doping dependence behind the first CP. It
reaches the T$_c$ - dome. Model calculations have been made. The
peculiarities of the phase diagram are determined by the doping-stimulated
band overlap dynamics. The observation of ,,foreign`` carriers as
compared to the ones introduced by doping  is expected to start with the vanishing PG.

 \noindent PACS: 74.20.-z; 74.72.-h; 74.25.Dw
\end{abstract}

\begin{keyword}
Electron-doped cuprates; Superconductivity; Pseudogaps
\end{keyword}

\end{frontmatter}

\maketitle   
\section{Introduction}
The concept of the pseudogap has born from the analysis of hole-doped
cuprate superconductor low-energy excitations.
An extensive literature
exists concerning this feature and its interpretation [1-9].
The PG has been identified by various experimental methods and seems to be a
characteristic phenomenon of doped cuprate superconductors. It is nearly material-independent [10].
The PG is exposed also in the normal state material at T$>$T$_c$. This is in essential contrast with the
true strongly material dependent superconducting gap (SCG) observed at  T$<$T$_c$. There have been misunderstandings in
literature in discussing various gaps in comparable conditions. The PG can enter
the superconductivity dome and  be observable in this region as a normal state gap (NSG) [11].
Below  T$_c$ the PG can remain masked by the excitations of superconducting states. At a fixed doping
there  is no transformation of the PG into a SCG.

The magnitude of the PG is usually characterized by
a temperature  T$^*$ which is strongly doping-dependent. This T$^*$ line does not represent a real phase
transition [12, 13]. The PG cannot be directly jointed to the superconducting order through its fluctuations in the normal
state. One has shown that these fluctuations are quenched far below  T$^*$ [14-16].

At present the origin of the PG is attributed to extrinsic genesis in respect of superconductivity [2, 6, 14, 16-19].
The formation of the PG can be associated with various types of the orderings of lattice and electronic nature [20-25], accompanied
also with time reversal, spatial rotations and translation symmetry breaking [26, 27].

These orderings can be competitive with  superconductivity in distinct momentum regions which otherwise participate in
pairing. Then a depletion of excitation density around the Fermi energy appears.

The presence of a magnetic counterpart of the charge-channel PG has also been reported [1, 16, 28]. In general, the role of the spin-subsystem
remains in the PG aspect not well understood.

Both the PG and the SCG appear as quasiparticle-type excitations. However, the SCG is of coherent- and the PG of noncoherent
nature excitation [26, 29, 30].
These gaps  can  represent the actions of different momentum space regions at the same dopings.

The cuprate superconductivity is initiated by a necessary doping process. The first major result of this
procedure is the metallization of the sample. It means that the chemical potential ($\mu$) must lie in a partially filled band
of the doping influenced spectrum. The basic  (AF) order of cuprate becomes destroyed. The distortions
of the crystal interior, which create the PG in the spectrum of the doped material, appear. This process at very low dopings is
better known for hole-doped samples. It starts by forming hole-centered ferrons [31] on  AF networks [32, 33]. An extended
doping stimulates the percolation of such mesoscopic dopants-associated CuO$_2$ plane plaquettes [32]. A recent paper
[34] has followed on the atomic scale the  evolution of the electron structure from an insulator to superconductor,
passing some kind of weak insulator state.
The PG becomes built up on the background of the weakening of AF correlations
with an extended doping.
For electron-doped systems, where the AF and superconducting domains overlap
markedly, the problem needs further investigations.

The PG and the SCG can be observed in the same underdoped region. The energy scales for these two types of gaps are here rather different,
 as also their development with doping are opposite. The PG diminishes, the SCG grows with doping. This is the simplest indication to differentiate them.
Coming nearer to the critical point (CP) on the phase diagram, the mentioned mismatch in energy is vashed out.
The phase diagram CP is defined as the doping concentration,
where the NSG becomes closed [28, 35-37]. Fermi surface reconstruction appears here. Behind the CP the mixed carrier system behaves more as a Fermi
liquid.

The PG appears as being selective in respect of momentum space regions, where it is created and observed in the corresponding spectral windows. In hole-doped superconductors the PG is merely an ''antinodal'' ( ($\pi,0$)-type Brillouin zone centered regions) event. The SCG at underdoping
is of      ($\pi/2,\pi/2$)
type ''nodal'' phenomenon. On the route between these points the PG is depressed. The carriers in the regions where the PG is formed must behave
as insulating below T$^*$. When passing this curve insulator to metal transitions in the momentum-restricted region are expected and observed [38-40].

\section{The pseudogap on multiband background}

Cuprates are electronically highly correlated materials. The reconstruction of the crystal interior under doping must be taken into account.
The Mott-Hubbard type AF spectrum of a pure sample will be reorganized to form the superconducting playground.
This is the second essential result due to doping.
The perturbed spectrum includes a
qualitatively essential effect. New electron states (bands) inside the charge-transfer gap are created [26,41-46]. One speaks about the creation of midgap-, polaronic-,
or defect bands. Doping can regulate the extension of such states and bring them into overlap with the itinerant bands. The superconductivity
acquires the multiband nature.

The electron spectrum near the Fermi surface, being actual for the superconductivity, is different for the hole and electron dopings. In the hole case the region
near the top of the (mainly) oxygen band between the upper (UHB) and lower Hubbard (LHB) is involved. For electron doping it is the region near the bottom
of the UHB.

For  superconductivity the part of the elecron spectrum near the Fermi surface is acual.
The modifications due to doping with the  creation  of the new (defect)
bands appear just here. For hole and electron dopings these spectral regions are different [41, 42, 72].
As largely known, in the hole case it is the area near the top of the (mainly) oxygen band.
This band is positioned between the upper (UHB) and lower Hubbard (LHB) bands of the pure AF
sample. For the electron doping the region near the bottom of the UHB has been established to be active.

The cuprate superconductivity mechanism remains not well understood at present, c.f. e.g.
[47-51, 56]. The approach of our group has been based on a multiband nature of the doped
cuprate, with the original idea to use the interband pairing interaction between the itinerant and doping-induced bands. Model realizations of such type approaches for hole-doped cuprates have given general conclusions and results
in a qualitative agreement with experimental data [56,57]. The PG on the phase diagram appears here naturally [56-62]. The later
consideration of electron-doped systems by analogous interband pairing scheme can be followed in [56,62]. The
global  similarity of the phase diagrams for both types of doping  arise from the ''topologically''
comparable physical approach. The essential differences between them stem from the perturbation of different atomic sublattices  of the material.

Recently the pairing glue in the cuprates has been identified as being of electronic origin
[63]. This justifies the use of interband pairing interaction in our approach (exchange type pair transfer) with its outstadning advantages [64].

In this paper we analyse the PG-s in electron doped cuprates. This problem is not very clear - the recent review paper [65] summarizes that ''there are substantial signatures of PG effects
in the electron-doped compounds''. In this aspect also papers [3, 15, 66-68]
can be consulted and theoretical approaches [69-73] must be mentioned.

 Our original definition [55, 58, 59] of  PG excitations  exposes the PG as a natural phenomenon on the multiband background [74].
It is based on the possibility of the appearance of bands intersected or not intersected by the chemical potential ($\mu$). Namely, the PG is considered as the minimal quasiparticle
excitation energy of a band ($\alpha$) not containing $\mu$
\begin{equation}
\label{eq1}
\Delta_{p \alpha } =\left\{ {\vert \xi_\alpha -\mu \vert_{\min }^2
+\Delta_\alpha^2} \right\}^{1/2}.
\end{equation}
It is supposed that $|\xi_\alpha-\mu|_{min} \neq 0$.
In (1) $\xi_\alpha $ stands for the band energy and $\Delta_\alpha $ for the
SCG gap of the same band.
In the case of a nonrigid spectrum $\mu$ can be shifted into $\xi_\alpha$, so that $|\xi_\alpha-\mu|_{min} = 0$.
Then the normal state contribution into $\Delta_{p \alpha }$  vanishes.
One reaches a phase diagram CP.
Behind a CP the $\Delta_{p \alpha }  $ remains represented by the sole SCG $\Delta_\alpha
$.  The reorganizations in the multiband spectrum by doping, or, say, by pressure can influence the PG until its disappearance.
The changes in band overlaps become of first significance in the problem of the PG. When the normal state contribution prevails,
the PG will be large and repeats practically the behaviour of the NSG. In the neighbourhood of the CP $\mu$ enters the given band (or the band reaches $\mu$).
The normal state contribution into the PG will be small and both components of $\Delta_{p \alpha } $ can show a comparable magnitude. The quenching
of superconductivity ($\Delta_{\alpha } $) can then extract the NSG in this spectral region.

\section{Multiband model for an electron-doped cuprate}

Basing on versatile knowledge, a 4-band model of the active part of electron-doped cuprate spectrum
(Fig.~1) has been proposed [62].
There is the UHB --($\beta )$ which extends up
from its antinodal (AN) ($\pi ,0)$-type bottom at $d_1 $. The LHB-($\gamma )$ is
situated deep and has a nodal (N) $\left( {\frac{\pi }{2},\frac{\pi }{2}}
\right)$ -type top. Both these bands are of width $\Delta d$. Midgap
 bands appear under the $\beta $-band bottom on the account
of both  Hubbard bands. To differentiate the functionality of the defect states in the momentum space we introduce two
subbands. The $\alpha_1 $ one is AN centered with the bottom lying at d.
The $\alpha_2 $ one is N-centered with the bottom taken as energy zero.
These bands arise from the perturbed copper sublattice and develop with the
destruction of the AF order, cf. also the recent calculation [25]. The total
number of states (4 bands) is normalized to 1. The concentration of doped
electrons for one Cu site per one spin is c. The undoped material (c=0)
has $\beta ,\gamma $-weights of states $\raise.5ex\hbox{$\scriptstyle
1$}\kern-.1em/ \kern-.15em\lower.25ex\hbox{$\scriptstyle 2$} $.
Electron-doping exhausts them to (1/2-c). The $\alpha$-bands are both of weight c.
There is a special spectral weight transfer from the Hubbard components to doping-induced midgap states.

Extended doping shifts the midgap bands towards the $\beta $-band bottom and
the bare gaps between them become removed. This well-known trend can be taken
into account by writing for the top energies of the $\alpha_1 $- and
$\alpha_2 $-bands $d+\alpha c^2$ and $\alpha c^2$ correspondingly. This
spectral behaviour is decisive for the behaviour of the PG-s on the phase
diagram. As a result, the progressive doping leads to the formation of a common
carrier family. The constant densities of states read $\rho_\alpha (1,2)=(\alpha c)^{-1}$;
$\rho_\beta =(\frac{1}{2}-c)(\Delta d)^{-1}$, i.e. the midgap bands grow
while the Hubbard spectrum tends to total ruination at $c=1/2$.

The chemical potential is shifted upward with electron doping starting from
a position inside the midgap states as expected, e.g. [65]. One finds the first
CP on this route to be reached at $c_0 =(d_1 -d)^{1/2}\alpha
^{-1/2}$. Here the first defect-UHB ($\alpha_1 -\beta $) gap becomes closed. One finds for
$c\le c_0 $
\begin{equation}
\label{eq2}
\mu_1 =d+\alpha c^2,
\end{equation}
which lies on the higher midgap subband top ($\alpha_2 $ remains
filled). It is known [15] that at low dopings the electrons occupy ($\pi
,0)$-neighbourhood and extending Fermi surface pockets are formed. For $c\ge
c_0 $ the  upper midgap band and the UHB overlap and define a common
($c\ge c_0 )$
\begin{equation}
\label{eq3}
\mu_2 =(d+\alpha c^2+d_1 \alpha c\rho_\beta )(1+\alpha c\rho_\beta
)^{-1}.
\end{equation}
The second CP (c$_{x})$ appears when the nodal midgap states
reach the $\alpha_1-\beta$ mixture. Then the disposition of bands and $\mu $
tend to maximize T$_{c}$. So, at $c>c_x $ where $c_x =\sqrt {\mu_2 (c_x
)\alpha^{-1}} $,
\begin{equation}
\label{eq4}
\mu_3 =(d+2\alpha c^2+d_1 \alpha c\rho_\beta )(2+\alpha c\rho_\beta
)^{-1}.
\end{equation}
One really observes that now new spectral intensity around $\left(
{\frac{\pi }{2},\frac{\pi }{2}} \right)$ emerges to the game [75,78]. This
can be attributed to hole-like Fermi surface pockets formation [78]. The
joint smooth  rising up $\mu $-curve can be found in [62]. Note that the positions of the
phase diagram CP-s are essentially determined by the changes in the bands
overlap. The same concerns the natural
reconstruction of the Fermi surface with doping.


\section{ Two pseudogaps by electron doping}

There are two PG-s in our model connected to the bands $\beta $ (UHB) and the defect $\alpha_2 $
not intersected by $\mu $ at low dopings $c<c_0 $. The antinodal UHB PG
vanishes as the first CP is reached. At the same time, the
nodal-centered PG has a constant contribution corresponding to the energy
difference of both $\alpha $-bands. One has for $c<c_0 $
\begin{equation}
\label{eq5}
\Delta_{p\beta } =\left[ {(d_1 -d-\alpha c^2)^2+\Delta_\beta^2 }
\right]^{1/2},
\end{equation}
\begin{equation}
\label{eq6}
\Delta_{p\alpha 2} =\left[ {d^2+\Delta_\alpha^2 } \right]^{1/2}.
\end{equation}
At $c_0 $ the energy separation of occupied and empty states is removed.
This points to the vanishing long-range AF order. The calculation of the
magnetic susceptibility in a comparable case has shown [75] that deviations
from the half-filling cause rapid reduction of the spin correlation length.

For $c>c_0 $ the chemical potential becomes changed to $\mu_2 $(Eq.3) as
$\alpha_1 $- and $\beta $-bands overlap. The smaller PG exhibits now
an exposed dependence on doping
\begin{equation}
\label{eq7}
\Delta_{p\alpha 2} =\left[ {(\mu_2 -\alpha c^2)^2+\Delta_\alpha^2 }
\right]^{1/2}.
\end{equation}
This gap closes at $c_x $ in the normal state.
There remain no NSG near the UHB bottom.
One expects that here the
tracks of the (short-range) AF correlations, which survived until $c_x $ will be
quenched. The AF order vanishes now at all. As a result, two PG-s, having
their own relation to the AF order, are expected for electron-doped cuprate
superconductors. However, the different assignment of ($\alpha_1; \alpha_2$)
generic momentum space points must be held in mind.
The presence of two PG-s in R$_{2-x}$Ce$_x$CuO$_4$ seems to follow from the data of [66],
see also [65].

The appearance of two PG-s in our model is the result of the supposition that the defect ($\pi, 0$) and  ($\pi /2, \pi /2$)
centered regions of the momentum space are created with a gap between them. Without this assumption there would be only one SCG and one PG present.

\section{ Results}

The mean field Hamiltonian for the proposed 3-band model of the active part includes the UHB and
$\alpha_1$, $\alpha_2$ defect bands.

$$
H=\sum_{\sigma ,\vec{k},s}\epsilon_{\sigma}(\vec{k}) a\sp
+_{\sigma ,\vec{k},s}a_{\sigma ,\vec{k},s}+ \Delta_{\beta}
\sum_{\vec{k}} [a_{\beta\vec{k}\uparrow}a_{\beta
-\vec{k}\downarrow}+ a\sp +_{\beta -\vec{k}\downarrow}a\sp
+_{\beta \vec{k}\uparrow}]
$$
\begin{equation}
-\Delta_{\alpha}\sum_{\vec{k},\alpha}{}\sp{1,2}
[a_{\alpha\vec{k}\uparrow}a_{\alpha -\vec{k}\downarrow}+ a\sp
+_{\alpha -\vec{k}\downarrow}a\sp +_{\alpha \vec{k}\uparrow}]\; .
\end{equation}

Here the band energies
$\epsilon_{\sigma}=\xi_{\sigma}-\mu$ are counted from the chemical potential.
The LHB remains filled and figures only in the determination of $\mu
$. The constant interaction strength ($W$) of the UHB with both defect bands are taken equal
(so $\Delta_{\alpha 1}=\Delta_{\alpha 2}$). The integration ($\vec{k}$) for defect bands
is performed over the correspoding energy intervals with their densities of states.

The Hamiltonian (1) can be simply diagonalized and the results for superconducting gaps with their  definition
 are given by the following expressions ($\Theta =k_B T $)

\begin{eqnarray}
 \Delta_{\beta}=2W\sum_{\vec{k},\alpha}{}\sp{1,2}\langle
a_{\alpha\vec{k}\uparrow}a_{\alpha -\vec{k}\downarrow}\rangle \ =W\Delta_{\alpha}\sum_{\vec{k},\alpha}{}\sp{1,2}
E_{\alpha}\sp{-1}(\vec{k}) th\frac{E_{\alpha}(\vec{k})}{2\Theta} ,\\
\nonumber \Delta_{\alpha}=2W\sum_{\vec{k}}\langle a_{\beta
-\vec{k}\downarrow}a_{\beta \vec{k}\uparrow}\rangle =W\Delta_{\beta}\sum_{\vec{k}}
E_{\beta}\sp{-1}(\vec{k}) th\frac{E_{\beta}(\vec{k})}{2\Theta} \; .
\end{eqnarray}

The quasiparticle energies are of an usual form $
E_{\sigma}=\sqrt{\epsilon_{\sigma}\sp 2
+\Delta_{\sigma}\sp 2}. \;
$  These expressions are used   in writing (1) and (9). Gaps (9) tend to zero at a common $T_c$.

The calculated normal state pseudogaps and the T$_{c}$ domain for a
,,representative`` electron-doped cuprate are shown in Fig.2. The following
input parameters have been used: $d=0.05$; $d_1 =0.15$; $\Delta d=0.8$
and $\alpha=11 $(eV). The interband coupling constant W=0.17 eV
between the midgap bands and the upper Hubbard component guarantees then the
maximum $T_{cm} =31$K at $c=0.16$. The critical points of the model lie at
$c_0 =0.095$ and $c_x =0.15$. Reasonable modifications in this parameter set
do not lead to essential changes in the general results, cf. [62].

The first larger UHB-associated PG $\Delta_{
p \beta}$ falls to zero at $c_0$. This point is close to the onset of superconductivity.
The calculated [76] plane coherence length starts here a rapid drop, see also [75].
The behaviour of the spin stiffness can be followed in [77]. One expects near $c_0$ the long-range
AF correlations being removed.

The second CP $c_x > c_0$ marks the disappearance of the smaller PG $\Delta_{p \alpha}$
and the vanishing of the survived short-range AF correlations as well, i.e. the AF correlations
altogether. This conclusion is to some extent supported by the finding that with approaching the superconducting order (and vanishing PG)
the AF correlations are of a short-range and not of the long-range nature [77]. Note also
that according to [78] the long-range AF order is not necessary for the development of the PG.

The second CP appears at a position  inside the superconducting dome before the maximum $T_c$ is reached at optimal doping.
This means that the domains, where (short-range) AF fluctuations are preserved, overlap the superconductivity region markedly.
This circumstance is usually mentioned [64], especially in
comparison with the hole-doped case. Beyond $c_x $ the low-energy
excitation spectrum is expected to be presented by the  superconducting
gaps $\Delta_\beta $ and $\Delta_\alpha $.
The large Fermi surface composed of all overlapping bands supports the
optimization of $T_c $.



When analysing the spectral manifestations of multiple pseudo- and superconducting gaps,
one must be careful. The smaller gap can remain hidden under the larger one and it can also mask
the sharpness of the larger gap.
There can be problems with the separation of the constant $\Delta_{ p \alpha} $ on the background
of $\Delta_{ p \beta} $   without considering the
angular polarization difference.

The superconducting contribution $\Delta_\alpha $ into $\Delta_{p\alpha }
$ becomes remarkable with an extending doping towards $c_x $. This is
illustrated in Fig.3.


The bands overlap dynamis of the present model allows one to explain some new
effects. This can be illustrated by  Fig.4, for example, for $\alpha_2 $ and
$\beta $ bands being shifted in the momentum space. Namely, the presence of hole-
and electron-type carriers has been detected in electron-doped
superconductors [64,76,77]. In the case of the $\mu $ position in Fig.4 it
intersects both the electron- and hole-like pockets. Analogously, the
observation of steep and flat dispersions for carriers [81] can be
explained. Also, one can expect the observation of the ,,foreign``-type
carriers as compared with the ones introduced by doping. However, there is an essential
requirement that the corresponding ,,active`` PG has been closed (upper band
in Fig.4).

In addition, paper [82] states that although both hole and electron
pockets are known in electron-doped cuprate superconductors, there ,,is no
evidence so far`` of electron pockets in hole-doped cuprates. This finding
can be explained by using the mentioned relative behaviour of the PG and the
,,foreign`` carriers effect. For electron-doped materials it is known,
as also found in the present work, that the PG (curve 3 in fig.2) touches the
$T_c $ domain at a relatively ,,early`` level. Contrary to the hole-doped
materials, the end of the PG is found (also calculated [56]) to extend nearly
until the overdoped end of the $T_c $ domain.
The doping interval for the effect under consideration will be narrow.
In general, the observation
 of "foreign" carriers as compared to the ones introduced by doping  is expected to
start with the vanishing of the corresponding PG.

An extended comparison of the present model, formulated for a ,,typical``
electron-doped cuprate superconductor, with the experimental data is a
complicated task. The mainly investigated systems are of the type
R$_{2-x}$M$_{x}$CuO$_4$ (R -- rare earth; M -- Ce or Th). The data for a
distinct material and (naturally) between various systems are dispersive. It
seems that the general results obtained by our model mechanism justify its
use in various analyses.
Especially the rised problem of the existence of two PG-s in electron-doped cuprates is of interest.
The theoretical approach itself can be improved in
various aspects, e.g.  taking  the pairing strengths for the $\beta -\alpha
1$ and $\beta -\alpha 2$ channels to be different, or precising the parameter
set.

\section{Conclusion}

In conclusion, the present model approach suggests the appearance of two
pseudogaps in electron-doped cuprates. These gaps are connected with the
destruction of long- and short-range antiferromagnetic correlations. There
are two critical points on the doping phase diagram. The larger one is centered
in the antinodal momentum space region at small dopings. The second one is
expected to be mainly a near-nodal event. Its doping dependence is exposed
behind the first critical point. The second critical point lies before the
maximal $T_c $. Antiferromagnetic and superconducting domains overlap
markedly.

Nowadays  multiband pairing schemes are applied to
novel, famous or artificial superconducting systems. The
appearance of a PG-type excitation in the spectra of these
materials can be considered, as illustrated also in the
present work, to be an indication of a multicomponent
(probably three-band [61]) nature of the superconducting playground.
On the other hand, in the case of a multicomponent Fermi surface intersected simultaneously
by the chemical potential, one can expect high transition
temperatures by the functioning of interband coupling
channels (missing PG).

This work was supported by the European Union Regional Development Fund
(Centre of Excellence ,,Mesosystems: Theory and Applications``, TK114) and
the Estonian Science Foundation grant No 8991.

\textbf{References}

[1] T. Timusk and B. Statt, Rep. Prog. Phys. 62 (1991)
61.

[2] T. Timusk, Solid State Commun. 127 (2003) 337.

[3] A. Damascelli, Z. Hussain and Z.-X. Shen, Rev. Mod. Phys. 75 (2003) 473.

[4] D. N. Basov and T. Timusk,
Rev. Mod. Phys. 77, (2005) 721.

[5] P. A. Lee, N. Nagaosa and X. G. Wen, Rev. Mod. Phys.  78 (2006) 18.

[6] J. L Tallon et al., Physica C 415 (2004) 9.

[7] S. H\"{u}fner et al., Pep. Prog. Phys. 71 (2008) 062501.

[8] R. Daou et al., Nature 463 (2010) 519.

[9] T. M.  Rice, K.-Y. Yang and F. C. Zhang, Rep. Prog. Phys. 75 (2012) 016502.

[10] T. Yoshida et al., J. Phys. Soc. Jpn. 81  (2012) 011006.

[11] S. H. Naqib et al., Phys. Rev. B 71  (2005) 054502.

[12]  L. P. Gor'kov and G. B. Teitel'baum,  Phys. Rev. Lett. 97  (2006) 247003.

[13]  L. Yu et al.,  Phys. Rev. Lett. 100  (2008) 177004.

[14]  T. Kondo et al., Nature  Phys. 7  (2011) 21.

[15]  K. Okada,  J. Phys. Soc. Jpn. 78  (2009) 034725.

[16]  N. Bergeal et al., Nature Phys. 4  (2008) 608.

[17]  R. Khasanov et al.,  Phys. Rev. Lett. 101  (2008) 227002.

[18]  J.-H. Ma, et al.,  Phys. Rev. Lett. 101  (2008) 207002.

[19]  A. Pushp, et al.,  Science 324  (2009) 1689.

[20]  A. Bianconi et al.,  Solid. State. Commun. 102 (1997) 369.

[21]  A. M.  Gabovich  et al.,  Adv. Cond. Mat. Phys. 2010 (2010) 68170.

[22]  F.  Lalibert\'{e}  et al.,  Nature Commun. 2 (2011) 432.

[23] C. V. Parker et al., Nature 468 (2010) 677.

[24] C. M. Varma and L. Zhu, Phys. Rev. Lett. 98 (2007) 177004.

[25]  V. Seagnoli et al., Science 332 (2011) 696.

[26] M. J. Lawler et al., Nature 466 (2010) 347.

[27] M. Hashimoto et al., Nature Phys. 6 (2010) 414.

[28] J. L. Tallon and J. N. Loram, Physica C 349 (2001) 53.

[29] J. Tahir-Kheli and W. A. Goddard, Phys. Chem. Lett. 2 (2011) 2326.

[30]  S. K. Nair et al., Phys. Rev. B 82 (2010) 212503.

[31]  V. D. Lakhno and E. L. Nagaev, Fiz. Tverd. Tela. 18 (1976) 3429.

[32] V. Hizhnyakov, N. Kristoffel and E. Sigmund, Physica C 160 (1989) 435.

[33] R. K. Kremer et al., Zs. f. Physik B 86 (1992) 319.

[34] Y. Kohsaka et al., Nature Phys. 8 (2012) 534.

[35] C. Varma, Nature 468 (2010) 184.

[36] L. Taillefer, Annu. Rev. Condens. Matter Phys. 1 (2010) 51.

[37] D. Di Castro et al., Eur. Phys. J. B 18 (2000) 617.

[38] P. Calvani, Phys. Stat. Solidi b 237 (2003) 194.

[39] F. Venturini et al., Phys. Rev. Lett. 89 (2002) 107003.

[40] P. Fournier et al., Phys. Rev. Lett. 81 (1998) 4720.

[41] T. Takahashi et al., Physica C 170 (1990) 416.

[42] C. Uchida et al., Phys. Rev. B 43 (1991) 7942.

[43]  M. B. J. Meinders, H. Eskes and G. A. Sawatzky, Phys. Rev. B 48 (1993) 3916.

[44] P. Schwaller et al., Eur. Phys. J. B 18 (2000) 215.

[45] A. Bianconi et al., Physica C 296 (1998) 269.

[46] A. Ino et al., Phys. Rev. B 65 (2002) 094504.

[47] A. S. Alexandrov, J. Supercond. 18 (2005) 3.

[48] R. Micnas, S. Robaszkiewicz and A. Bussmann-Holder, Physica C 387 (2003) 58.

[49] H. Keller and A. Busmann-Holder, Adv. in Cond. Matter Phys. ID393526 (2010).

[50] D. Innocenti et al., Phys. Rev. B 82 (2010) 184528.

[51] H. Kamimura et al., Theory of Copper Oxide Superconductors, Springer, Berlin, 2005.

[52] P. Konsin, N. Kristoffel and T. \"{O}rd, Phys. Lett. A 129 (1988) 339.

[53] N. Kristoffel, P. Konsin and  T. \"{O}rd, Rivista Nuovo Cim. 17 (1994) 1.

[54] N. Kristoffel, Phys. Stat. Solidi b 210 (1998) 195.

[55] N. Kristoffel and P. Rubin, Physica C 356 (2001) 171.

[56] N. Kristoffel, P. Rubin and  T. \"{O}rd, Intern. J Modern Phys. B 22 (2008) 5299.

[57] N. Kristoffel, P. Rubin and  T. \"{O}rd, J. Phys. Conf. Ser. 108 (2008) 012034.

[58] N. Kristoffel and P. Rubin, Solid State Commun. 122 (2002) 265.

[59] N. Kristoffel and P. Rubin, Physica C 402 (2004) 257.

[60] N. Kristoffel and P. Rubin, Phys. Lett. A 374 (2009) 70.

[61] N. Kristoffel and P. Rubin, Intern. J Modern Phys. B 26 (2012) 1250144.

[62]  N. Kristoffel and P. Rubin,  Phys. Lett. A 372 (2008) 930.

[63]  V. M. Krasnov, S.-O. Katterwe and A. Rydh, Nature Commun. 4 (2013) 3970.

[64]  N. Kristoffel: in Progress in Superconductivity Research, Ed. O. A. Chang, Npva Sci. Publ., NY, 2008, p.7.

[65] N. P. Armitage, P. Fournier and R. L. Greene, Rev. Mod. Phys. 82 (2010) 2421.

[66] L. Alff et al., Nature (Lett.) 422 (2003) 698.

[67] K. Jin et al., Nature (Lett.) 476 (2011) 73.

[68] S. Basak et al., Phys. Rev. B 85 (2012) 075104.

[69] Y. Onose et al., Phys. Rev. Lett. 87 (2001)  217001.

[70] C. Kusko et al., Phys. Rev. B 66 (2002) 140513(R).

[71] B. Kyung et al., Phys. Rev. Lett. 93 (2004) 147004.

[72] M. Taguchi et al., Phys. Rev. Lett. 95 (2005) 177002.

[73] C. Weber, K. Haule and G. Kotliar, Nature Phys. 6 (2010) 574.

[74] N. Kristoffel, in: J. E. Nolan (Ed.), Superconductivity Research Advances, Nova Sci. Publ. Inc., 2007, p.225.

[75] A. Sherman and M. Schreiber, Phys. Rev. B 76 (2007) 235112.

[76] N. Kristoffel and P. Rubin, in: Electron Transport in Nanosystems, Ed. J. Bon\v{c}a, S. Kruchinin, Springer Sci, Dordrech, 2008, p. 179.

[77] M. Motoyama et al., Nature 445 (2007) 186.

[78] B. Kyung et al. Phys. Rev. Lett., 93, (2004) 147004.

[79] H. Matsui et al., Phys. Rev. Lett. 94 (2005) 047005.

[80] Y. Dagan and  R. L. Greene, Phys. Rev. B 76 (2007) 24506.

[81] S. Park et al., Phys. Rev. B 75 (2007) 060501(R).

[82] S. Chakravarty, Rep. Prog. Phys. 74 (2011) 022501.

\textbf{Figure captions}

Figure 1. The bands scheme with doping-created midgap (defect) bands ($\alpha
_1 , \alpha_2 )$ between the LHB and UHB. O denotes the mainly oxygen band.
Nodal symbols belong to band tops.

Figure 2. The doping phase diagram of an electron-doped cuprate
superconductor with the $T_c $ dome and the larger (\ref{eq1}) and smaller (\ref{eq2})
normal state PG-s.

Figure 3. The behaviour of the PG $\Delta_{p_\alpha } $ (\ref{eq1}) near the
second critical point $c_x$. Behind $c_x $ it continues as $\Delta_\alpha $. For
$c<c_x $ the $\Delta_\alpha $ (2) is hidden in $\Delta_{p \alpha 2 }$. (3)- the normal
state contribution to the PG.

Figure 4. Bands crossing at different momentum points produce electron and
hole-pockets at the given $\mu $.
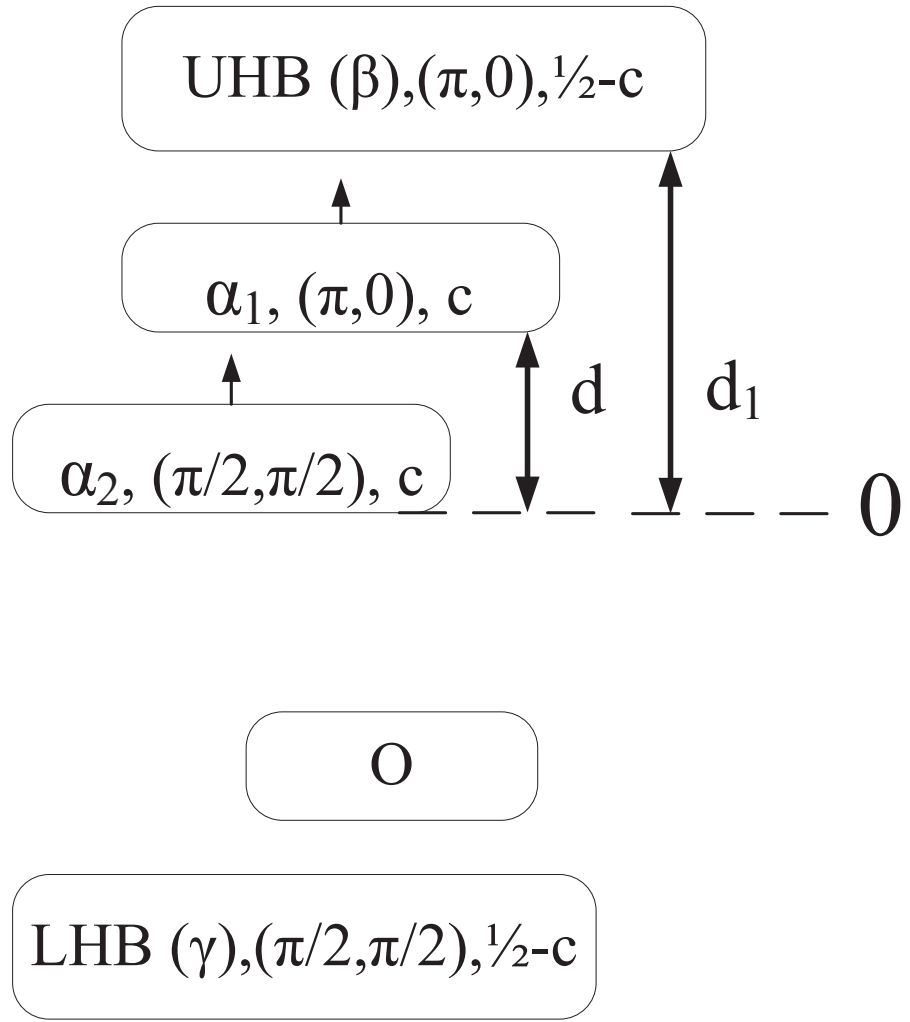
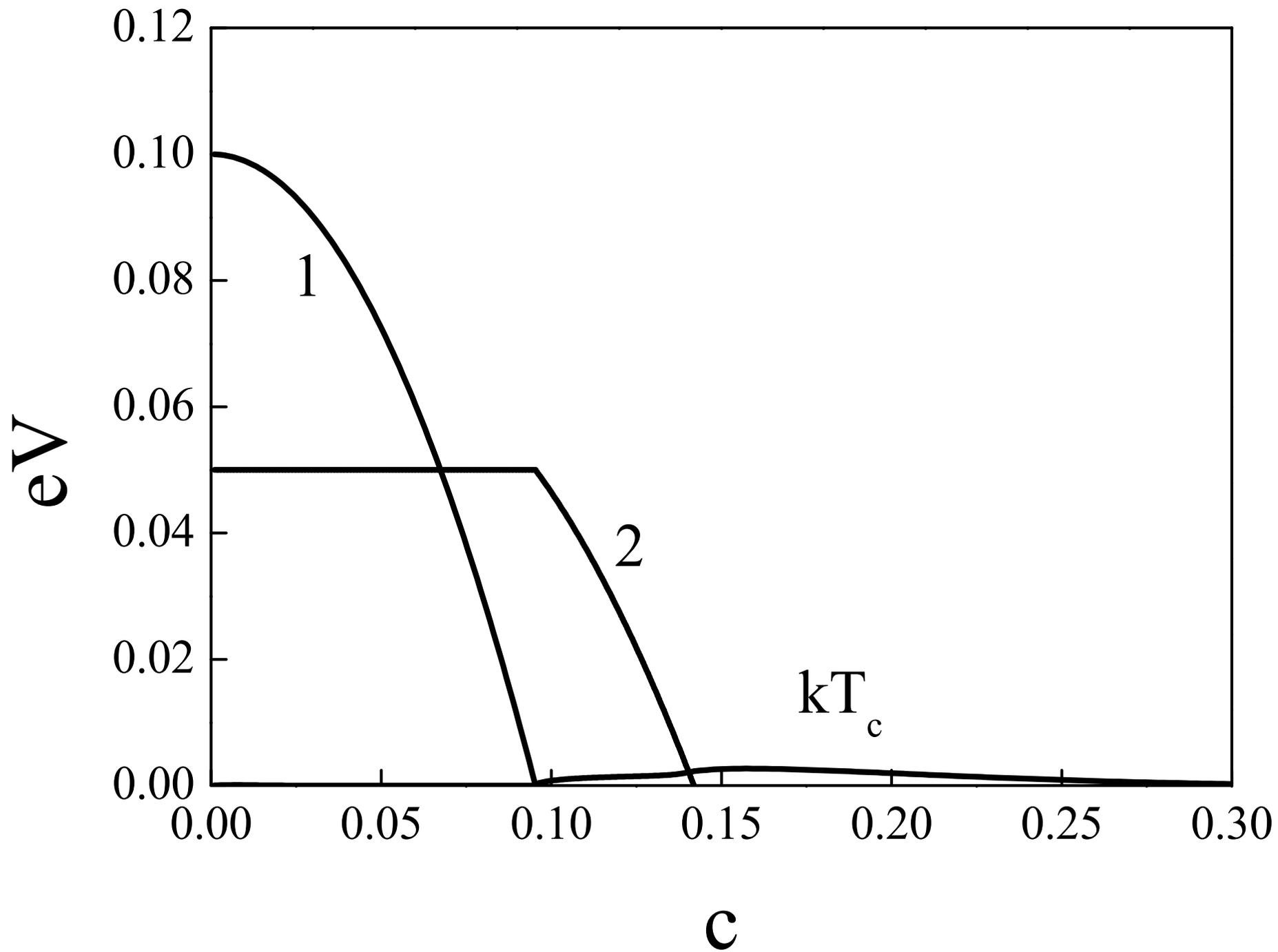
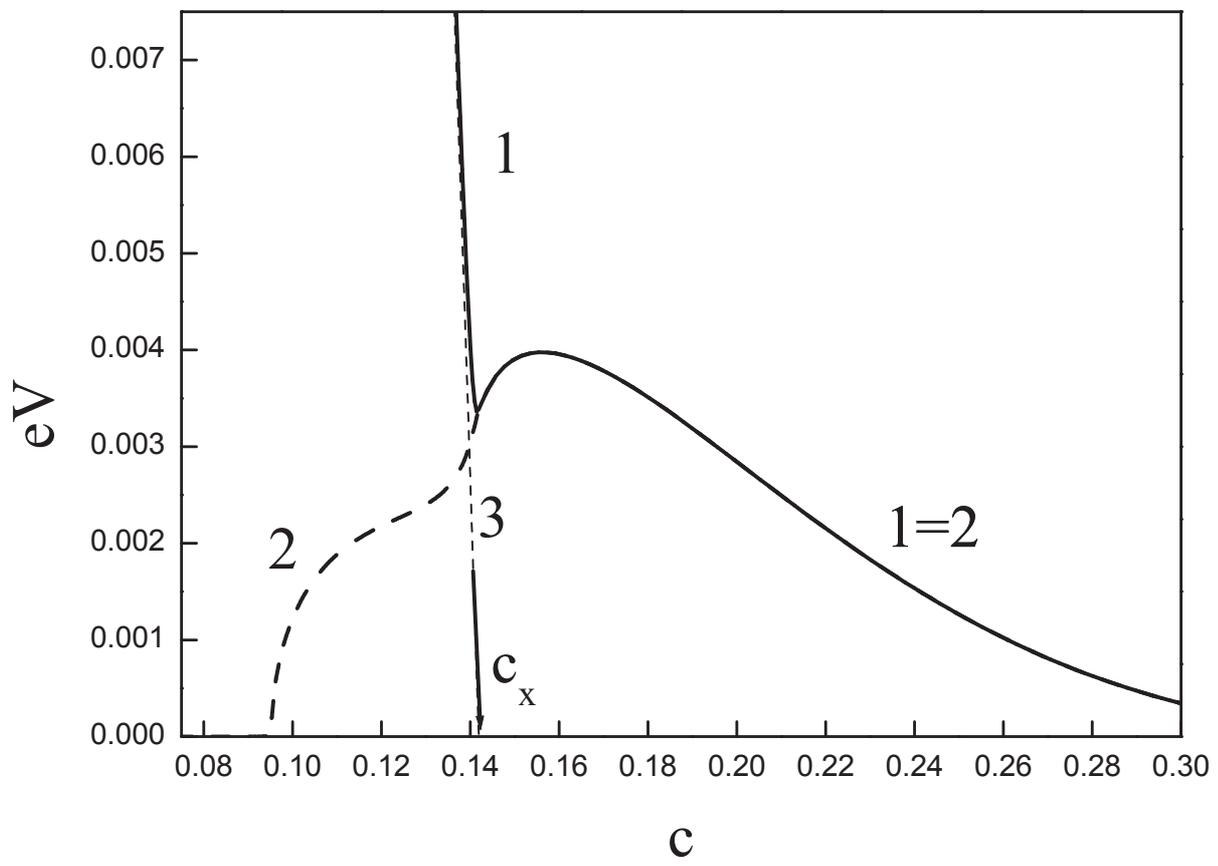
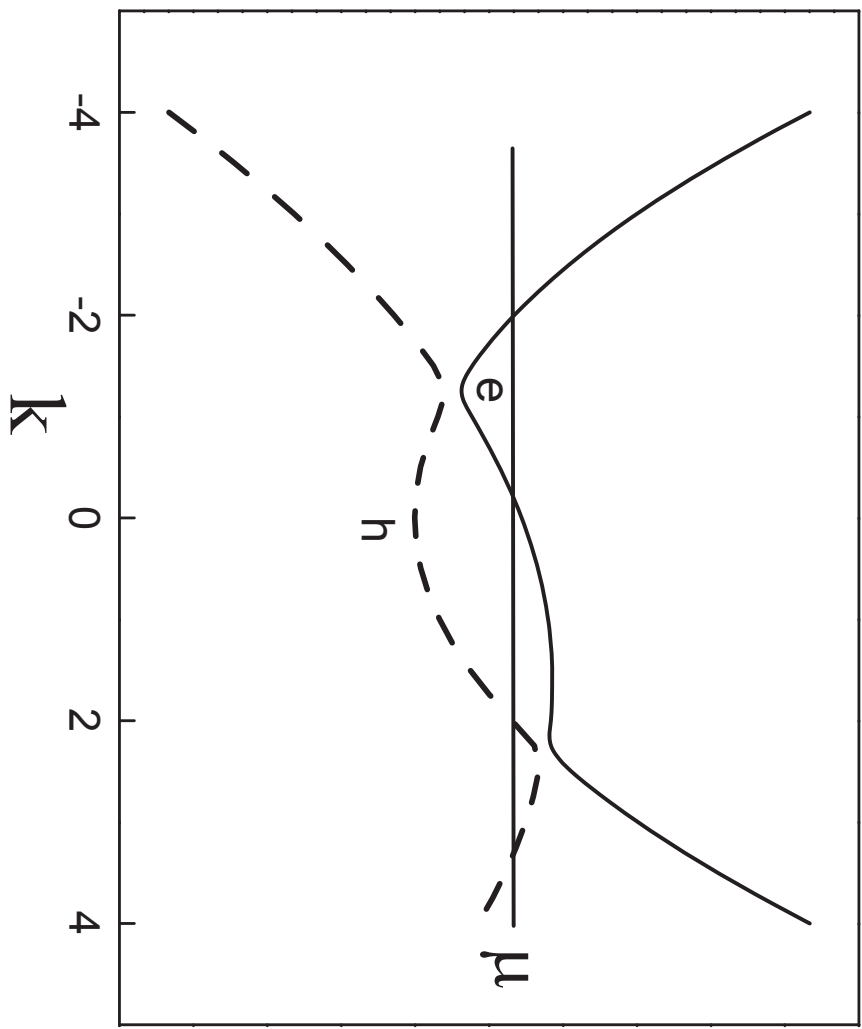
\end{document}